\documentstyle{mn}
\title{Reverberation by a relativistic accretion disk}

\author[Sergio Campana and Luigi Stella]
{Sergio Campana$^{1,\,2}$ and Luigi Stella$^{1,\,2}$\\
{$^1$ Osservatorio Astronomico di Brera, Via Brera 28, 20121 Milano,
Italy;}\\
{\hskip 0.33cm E-mail: campana@astmim.mi.astro.it and
stella@astmim.mi.astro.it}\\
{$^2$ affiliated to I.C.R.A.}}

\date{Accepted . Received , in original form }

\begin{document}
\newcommand{\be}{\begin{equation}}
\newcommand{\en}{\end{equation}}
\def\ltsima{$\; \buildrel < \over \sim \;$}
\def\lsim{\lower.5ex\hbox{\ltsima}}
\def\gtsima{$\; \buildrel > \over \sim \;$}
\def\gsim{\lower.5ex\hbox{\gtsima}}
\def\deg {^\circ}
\def\mdot {\dot M}
\def\kms {~km~s$^{-1}$}
\def\gs {~g~s$^{-1}$}
\def\ergs {~erg~s$^{-1}$}
\def\cmtre {~cm$^{-3}$}
\def\cmdue {~cm$^{-2}$}
\def\gcm {~g~cm$^{-3}$}
\def\rsole{~R_{\odot}}
\def\msole{~M_{\odot}}
\def\aa #1 #2  {{A\&A} {#1} #2}
\def\aas #1 #2  {{A\&AS} {#1} #2}
\def\araa #1 #2  {{ARA\&A} {#1} #2}
\def\mon #1 #2  {{MNRAS} {#1} #2}
\def\apj #1 #2  {{ApJ} {#1} #2}
\def\apjs #1 #2  {{ApJS} {#1} #2}
\def\astrj #1 #2  {{AJ} {#1} #2}
\def\nat #1 #2  {{Nat} {#1} #2}
\def\pasj #1 #2  {{PASJ} {#1} #2}
\def\pasp #1 #2  {{PASP} {#1} #2}
\def\ref{\par\noindent\hangindent=.8truecm}

\label{firstpage}

\maketitle

\begin{abstract}
We calculate the response of a line emitted from a relativistic accretion
disk to a continuum variation of the central illuminating source. This
model might be relevant to the very broad and sometimes redshifted
emission lines which have been observed at optical and X-ray
energies in Active Galactic Nuclei and X-ray binaries. We consider
separately the cases of a point-like and of an extended central source.
Based on the first three moments of the line energy distribution (i.e.
the line intensity, centroid energy and width) and the two-dimensional
transfer function, we identify a number of characteristic features,
which can help assessing the presence of a relativistic accretion disk
and estimating its parameters. In particular, by combining an absolute
time measurement of the maximum of line response function with a
relative measurement in units of light crossing time the mass of the
central object can be derived.
\end{abstract}


\section{INTRODUCTION}

It has long been suspected that the innermost regions of accretion flows
towards the central object in Active Galactic Nuclei (AGNs) might be
characterised by a disk geometry. The blue-UV bump (Malkan \& Sargent 1982)
and/or soft X-ray excess (Arnaud et al. 1985) observed in a number of quasars
and Seyfert galaxies, provide indirect evidence in favour of this
interpretation.
The size of the blue-UV emitting region is constrained by the
measurement of the time delay which characterises the response at
different wavelengths to variations of the ionising flux at the central
source.
{}From the monitoring of the Seyfert galaxy NGC~4151 (Clavel et al. 1990;
Ulrich et al. 1991) a time delay of $4\pm 3$~d between the UV-continuum
variations and the response of two different lines has been estimated. A
longer observing campaign for NGC~5548 (Clavel et al. 1991; Krolik et al.
1991) provided an upper limit of $\sim 4$~d to the delay between the
optical-UV continuum variations at four different wavelengths.
Reprocessing
from accretion disk matter of the primary (hard) X-ray radiation can
explain the absence of substantial delays between the optical and
ultraviolet light curves (e.g. for NGC~5548 Molendi, Maraschi \& Stella 1992).

The presence of a relativistic accretion disk can in principle be
established on the basis of the characteristic double-horned line
profile which arises from a flat Keplerian disk with a
power-law emissivity dependence (Chen \& Halpern 1989; Fabian et al. 1989).
Broad asymmetric double-horned Balmer ($H\alpha$, $H\beta$) emission
lines have been observed in several broad-line radio galaxies
and radio loud quasars (Eracleous \& Halpern 1994 and references therein).
However double-horned line profiles are rare in AGNs and, even in these
cases, some observations show features that
are inconsistent with the simple and stationary relativistic accretion
disk modelling (e.g. the red horn occasionally becomes brighter than the
blue horn in 3C390.3, Zheng, Veilleux \& Grandi 1991; see also Eracleous
et al. 1995).
{}From this discussion we can conclude that testing disk models
based only on stationary profiles can be difficult or inconclusive.

The way matter ``reverberates" the incident radiation from the central
source provides an additional diagnostic that can be used to infer the
geometry and the dynamics of the line emitting region (Blandford \& McKee
1982; Welsh \& Horne 1991; P\'erez, Robinson \& de la Fuente 1992).
The information derived from the stationary line profile and from
reverberation properties can therefore be compared for a consistency
check.
Based on the asymmetry and width of the stationary line profile,
Mannucci, Salvati \& Stanga (1992) estimate geometrical parameters and
subsequently derive the radial dependence of the line emissivity law by
adopting a deconvolution technique.
The geometrical parameters and the emissivity law are then used to derive
the expected response of the line intensity to continuum variations
to be compared with the existing observations.

Beside Balmer emission lines a promising feature to test the innermost
emission line regions in AGNs and X-ray binaries is likely represented by iron
K$\alpha$ lines (with rest energy of $6.4-6.9$~keV).
In a number of cases these lines are probably broader than $\sim 1$~keV
(FWHM) and/or redshifted to centroid energies as low as $\sim
6.0$~keV (e.g. Barr, White \& Page 1985; Day et al. 1990). Detailed line
profiles are not yet available as most studies have been carried
out with low resolution X-ray instrumentation ($E/\Delta E\lsim 10$)
and/or poor statistic (Smale et al. 1993).
The X-ray CCD detectors on board ASCA ($E/\Delta E\sim 50$) are
beginning to provide more detailed observations of iron K-shell lines
(Fabian et al. 1994).
In the most likely interpretation these broadened and shifted lines are
produced by bulk plasma motion in the vicinity of the accreting
collapsed object, possibly in a disk-like
geometry (Kallman \& White 1989; Fabian et al. 1989; George \&
Fabian 1991; Matt, Perola \& Piro 1991).
The re\-pro\-ces\-sing of hard radiation by cold matter in the form of an
accretion disk gives rise to iron K$\alpha$ lines, the centroid energy,
widths and equivalent widths (EWs) of which are in
agreement with the observed values (Nandra \& Pounds 1994).
Therefore iron K-shell lines might provide a sensitive diagnostic of
the innermost regions ($r\sim 10-100\,r_g$) of accretion flow towards
collapsed objects, where relativistic effects are expected to be important.
Detailed modelling of the physical conditions of a relativistic
accretion disk shows that more complex profiles than a simple double-horn
are expected in the case of high accretion rates (Matt, Fabian \& Ross 1993).

If reprocessing of hard X-ray radiation from a central source is
responsible for the production of iron K$\alpha$ lines in the innermost
region of accretion flows towards collapsed objects, then
reverberation studies relating line to continuum variations
can provide crucial information on the geometry and dynamics of the
line emitting region (Fabian et al. 1989).
As shown by Stella (1990), sufficiently short
variations at the central source should  give rise to two characteristic sharp
features in the profile that drift from the line wings towards the centre.
Such features provide in principle a powerful diagnostic of the
disk parameters and the mass of the central object.
In the absence of a detailed knowledge of the line profile,
the response of the line moments, in particular the
intensity, $I_l$, centroid energy, $E_c$, and width, $W$, can nevertheless
yield a wealth of useful information. Along these lines
Stella \& Campana (1991) and Campana \& Stella (1993) discussed a method for
constraining the disk parameters and the mass of the central object by using
the stationary line width and centroid energy of iron K$\alpha$ lines and
the cross-correlation delays between the continuum and the line intensity
variations.
Matt \& Perola (1992) first considered the detailed response of the iron
K$\alpha$ line centroid energy, intensity and width to continuum variations
from a point-like source above the disk plane. This geometry was introduced
in order to approximate an extended source and maximise the flux intercepted
by the disk.

In this paper we further investigate the response of a line from a
relativistic disk to variations at the central source,
by using the transfer function formalism (see Section 2). In Section 3 we
derive the line intensity, centroid energy and width for a flat, thin
accretion disk around a Schwarzschild black hole, taking into account
relativistic effects. We adopt two limiting geometries for the central
illuminating source: a point-like source lying in the disk plane and a more
realistic model involving a bulge-like source straddling the inner edge of
the accretion disk. For the first case we calculate also the two-dimensional
transfer function characterising the disk response.
Based on line moments, we identify and discuss several features that can be
used as indicators of geometry and scale length in a fashion similar
to the work of Matt \& Perola (1992). Our results are discussed in Section 4.

\section{REVERBERATION}

The response of a matter distribution to radiation from a central source
can be described with the formalism of transfer functions which relate the
line output to the intensity continuum input (Blandford \& McKee 1982).
As customary, we shall assume that the line emissivity is proportional
to the incident continuum flux.

At a given time $t$, the observer sees the line profile built up by
contributions of different portions of the line emitting region, each
characterised by an energy shift $1+z=E_0/E$, where $E_0$ is the line rest
energy.  Each portion is also characterised by a given light propagation
delay with respect to the observer, which regulates its response to
variations at the central source.
The response can, in general, be described through the two-dimensional transfer
function (2D-TF) $\psi(E,\,t)$, defined by:
$$
I_l(E,\,t)=\psi(E,\,t)*I_c(t)=\int\limits_{0}^{\infty}\psi(E,\,t-t')
\,I_c(t')\,dt'\eqno(1)
$$
with $I_l(E,\,t)$ the line intensity per unit energy seen by the
observer at a time $t$ and $I_c(t)$ the continuum
intensity; ``$*$" denotes a convolution.
The 2D-TF is the time-energy Green's function of the system.

The deconvolution of Eq.$\,$1 based on Fourier transforms requires high
signal-to-noise data and a nearby equispaced grid of observation times.
Two-dimensional maps of non-relativistic accretion disks
have been reconstructed through observations of emission lines in
cataclysmic variables, by using
the technique developed by Horne \& Marsh (Horne 1985; Marsh \&
Horne 1988; Marsh \& Horne 1990).
These authors apply a maximum entropy deconvolution technique to overcome
practical limitations due to relatively high noise and incomplete data sets
(Skilling \& Bryan 1984).
Observations are often too short and noisy to reconstruct
the 2D-TF, so it is useful to introduce the moments:
$$
I_l^{(n)}(t)=\int\limits_{-\infty}^{\infty}E^n\,I_l(E,\,t)\,dE
\hskip 1 cm \hbox { and }
$$
$$
\hskip 2cm
\psi^{(n)}(t)=\int\limits_{-\infty}^{\infty}E^n\,\psi(E,\,t)\,dE
$$
In particular, the one-dimensional transfer function (1D-TF) $\psi(t)\equiv
\psi^{(0)}(t)$ is:
$$
I_l(t)=\psi(t)*I_c(t)=\int\limits_{0}^{\infty}\psi(t-t')\,I_c(t')\,
dt'\eqno(2)
$$
The 1D-TF represents the line intensity $I_l(t)\equiv {I_l}^{(0)}(t)$
produced in
response to a $\delta-$fun\-ction continuum variation at $t=0$, i.e. it is the
time Green's function of the system.

Higher line moments can be derived in a similar manner; the line centroid
energy for a $\delta-$function continuum variation, normalised to the rest
line energy, is $E_c(t)=\psi^{(1)}(t)/\psi(t)$.
For the line width we use $W(t)=2.35\, \sqrt{[\psi^{(2)}(t)-E_c^2(t)]/
\psi(t)}$,
where a normalisation of $2.35$ is taken to
reproduce the FWHM in the case of a Gaussian profile.

Given $I_l(t)$ and $I_c(t)$ the 1D-TF can be derived from the deconvolution
of Eq.$\,$2. Blandford \& McKee (1982) apply this ``reverberation mapping
technique" to a variety of geometry and velocity distributions of
astrophysical relevance.
The 1D-TF for the $H\alpha$ line in NGC~4151 relative to the optical continuum
has been evaluated in this way (Maoz et al. 1991).
Maximum entropy deconvolution techniques have been used to derive
the 1D-TF for several emission lines (e.g. $Ly\alpha$, C~III] or $H\beta$) in
NGC~5548 relative to the optical and UV continuum (Horne, Welsh \& Peterson
1991), as well as the 1D-TF for $H\beta$ line in Mrk~590 relative to the
optical continuum (Peterson 1993).

\section{RELATIVISTIC ACCRETION DISK}

We assume a flat, geometrically thin disk, orbiting a Schwarzschild
black hole. The line emitting disk extends from $r_i$ to $r_o$
(measured in units of the gravitational radius $r_g=G\,M/c^2$)
and is observed at an inclination $i$.
To calculate the energy shift of a photon emitted by an infinitesimal
surface element of the disk at $r_{em}$ and $\phi_{em}$ ($\phi$ is the
azimuthal angle from the line of nodes with $0\deg < \phi<180\deg$ on the
side nearer to the observer) we use the fully relativistic treatment
described in the appendix of Fabian et al. (1989).
The energy shift $1+z$ is obtained by the product of two terms: one
term depends only on the emission radius and represents the strong field
equivalent of the gravitational and transverse redshifts; this term becomes
dominant at small radii. Besides the radius, the other term depends on the
relative orientation of the velocity of the disk matter and on the
direction of emission of photons reaching the observer and it can produce
either a blueshift or a redshift. This term corresponds to the Doppler
shift.
The intensity and the profile of a line arising from the reprocessing of
the central source radiation, depends also on geometry of the
central source. In the following we consider two limiting cases: a
point-like central source lying in the disk plane and a spherical source
straddling the disk inner edge.

\subsection{Point-like limit for the central source}

We consider here a central source which is point-like, emits isotropically
and lies in the disk plane.
Even if the EW of lines originating from thin disk reprocessing of central
source radiation is formally zero, this geometry is usually adopted to
discuss lines arising from disk-like broad-line regions (e.g. Welsh \&
Horne 1991; P\'erez et al. 1992), the dimensions of which are much
larger than the dimensions of the central source.

We assume that the line emitting disk matter responds instantaneously to the
photons from the central source and that the line photons freely propagate
away from the disk. The specific line intensity of the disk is approximated
by a $\delta-$function, i.e. $I_{E_{em}}=\epsilon(r,\phi) \> \delta (E_{em} -
E_0)$, where $E_0$ is the rest energy of the line and $\epsilon(r,\phi)$ the
surface line emissivity of the disk; phenomena affecting the local line
shape, such as line thermal broadening, blending or Compton scattering are
therefore neglected.
In stationary conditions, the radial dependence of the line emissivity
can be derived self-consistently through numerical integration once a
geometry for the central source and the disk is prescribed (Chen \& Halpern
1989; George \& Fabian 1991; Matt et al. 1991);
in the limit of a point-like central source (i.e. small respect to
the inner radius of the disk) lying in the disk plane,
the line emissivity varies as $r^{-3}$ (Mardaljevic, Raine \& Walsh 1988).

Flux variations at the central source induce variations in the line
emissivity with the delays introduced by light travel time effects.
Neglecting general relativistic corrections, the flux emitted at $r_0=0$ and
$\tau_0=0$ will be echoed, after a time delay $\tau$, by the disk region
defined by:
$$
\tau - \tau_0 ={{r-r_0}\over c} \,(1-\sin i \sin \phi) \eqno(3)
$$
Therefore, at any given time the echo front describes an ellipse
with eccentricity $\sin i$ and the central source in the most distant focus.
Moreover, the angular dependence of the echo front introduces
a $\phi$ dependence of the line surface emissivity.

Some relevant delays can be identified, which characterise the propagation
of the echo front on the disk: $\tau_1= r_i /c\, (1-\sin i)$,
corresponding to the echo front entering the line emitting portion of
the disk in the direction closest to the observer
(i.e. $r=r_i$ and $\phi=90\deg$);
$\tau_2=r_i/c$, the delay at which the echo front reaches
the line of nodes (i.e. $\phi=0\deg$ and $\phi=180\deg$) for $r=r_i$;
and $\tau_3=r_i /c\,(1+\sin i)$, the delay at which the entire echo front is
included in the line emitting region (i.e. $r=r_i$ and $\phi=270\deg$).

To calculate numerically the TFs and the line moments we adopt a short,
rectangular
impulse from the central source ($\Delta t=r_i/50\, c \ \ll r_i/c$), in
order to approximate a $\delta-$function
variation\footnote
{Note however that, from a physical point of view,
the shortest global variation at the central source,
in absence of other effects, should be $\gsim 2\,r_g/c$.
In this context we are interested in the
TFs and therefore we adopt a shorter variation.}.
With these assumptions, we compute the line profile produced in response by
a disk for a range of times and for selected values of the inner radius
$r_i$ and inclination $i$. In order to investigate the strong and weak
gravitational field regimes we adopt $r_i=6\,r_g$ (the radius of the
marginally stable orbit for a Schwarzschild black hole) and $r_i=50\,r_g$,
respectively. For the outer radius we use $r_o={10}^4 r_g$.

\subsubsection{Line moments evolution}

The line intensity response, $I_l(t)$, is shown in Fig.$\,$1.
For low inclinations ($i<30\deg$), the line intensity
shows a single peak centred around a time delay of $\sim 7-9\, r_g/c$ for
$r_i=6\, r_g$ (Fig.$\,$1a) and $\sim 60-80\, r_g/c$ for $r_i=50\, r_g$
(Fig.$\,$1b).
For higher inclinations an additional peak forms, which reaches its
maximum for time delays shorter than $r_i/c$.
For $r_i=50\, r_g$ the peak at short delays becomes dominant for $i\gsim
50\deg$ (note that for a classical disk this happens for $i\gsim 40\deg$). On
the contrary for $r_i=6\, r_g$ the peak at longer delays ($\tau_3$) always
dominates, because the gravitational and transverse redshifts decrease the
line intensity for small radii.
We note that in the case of a point source above the disk plane the relative
importance of the first peak is much reduced due to the different geometry
echo front (Matt \& Perola 1992).

Concerning the evolution of the line centroid energy, shortly after the echo
front has entered the disk ($\tau> \tau_1$), a redshift is produced by
gravitational and transverse shifts, which is more pronounced for smaller
inner radii (see Fig.$\,$2).
For low values of $i$ ($\lsim 40\deg$) the line centroid remains redshifted
also for larger time delays, due to the small projected velocities. On the
contrary, for higher values of $i$ a blueshift appears for time delays of
order $\tau_2$, due to the increasing importance of Doppler boosting.  The
blueshift reaches its maximum shortly after
the time when the echo front goes beyond the line of nodes ($\tau_2$).
In a weak field approximation, it can be shown that the largest shifts in the
echo front are produced for:
$$
\sin \phi_M = {1\over3} \Bigl({1\over{\sin i}}-\sqrt{\bigl({1\over{
\sin^2 i}}+3\bigr)}\Bigr)\eqno(4)
$$
i.e. regions $\lsim 20\deg$ beyond the line of nodes (Stella 1990).
The corresponding time $\tau'_2$, obtained by substituting $\sin\phi_M$ in
Eq.$\,$3, predicts quite accurately the maxima of the line
centroid energy for $r_i=50\,r_g$ (Fig.$\,$2b); due to more pronounced
photon deflection part of the accuracy
is lost in the strong field case ($r_i=6\,r_g$; Fig.$\,$2a).
As the echo front moves further outwards ($\tau\ge \tau_3$), the line
centroid energy approaches unity for any inclination,
due to the decreasing disk velocities and gravitational field.

In the weak field case ($r_i=50\,r_g$), the line width rises after
$\tau_1$, reaches a maximum at $\tau'_2$ when the highest redshifts and
blueshifts are also produced, and decays showing a flattening after $\tau_3$,
when the entire echo front is within the line emitting disk (Fig.$\,$3b).

The strong gravitational field case presents instead with a significant
difference (Fig.$\,$3a): the largest widths are achieved close to the
delay for which the entire echo front enters the disk (i.e. $\tau_3$).
This happens for any inclination.
General relativistic effects are responsible for the maximum around
$\tau_3$; due to the strong light bending in the innermost disk regions
($\sim 6\,r_g$), photons reaching the observer from $\phi\sim 270\deg$
are emitted locally at relatively small angles from the perpendicular to
the disk. Surface projection and gravitational lensing effects
therefore increase the flux from
these regions, which are also characterised by a range of shifts (see
also Matt, Perola \& Stella 1993; Rouchi \& Blandford 1994).
For inclinations $i\gsim 60\deg$, the relative importance of the
regions which, according to Eq.$\,$4, produce the highest shifts,
increases and a local maximum is produced for delays $\sim \tau'_2$.
We note that the evolution of the line centroid energy and line width are
remarkably similar to the ones derived by Matt \& Perola (1992) adopting a
point-like central source above the disk plane.

\subsubsection{2D-Transfer Function}

By definition, the two-dimensional transfer function, $\psi (E,\,t)$,
contains all the information on the evolution of the line profile and
higher line moments and provides in principle the most powerful diagnostic
of the line emitting disk region. Welsh \& Horne (1991) and P\'erez et al.
(1992) calculated 2D-TFs for different geometries in the non-relativistic
regime.
In particular, they calculate the 2D-TF for a Keplerian disk in the
non-relativistic regime and for a point-like central source lying in the
disk plane: in the time delay-energy plane the 2D-TF has a symmetric bell
shape (see Fig.$\,$5 in Welsh \& Horne 1991 and Fig.$\,$6 in P\'erez et al.
1992).
The 1D-TF as well as higher line moments can be easily recovered
from the 2D-TF by carrying out the relevant integration.
(Integration of the 2D-TF over the time delay axis provides the
stationary line profile).

Here we generalise these calculations in order to include general
relativistic effects by using the approximations of the disk model
described in Sect.$\,$3.
As expected the relativistic 2D-TF is strongly affected by relativistic
effects for small delays and gradually approaches the classical limit
as the echo front expands.
Fig.$\,$4a shows the 2D-TF for $i=10\deg$ and $r_i=6\,r_g$.
Because of the low disk inclination the projected velocities are small and
the line photons are strongly redshifted: these effects produce a narrow
2D-TF at small delays. For higher delays regions with larger velocities are
reached by the echo front and the 2D-TF rises monotonically approaching a
profile centered around unity. It is apparent that the centroid of the profile
grows monotonically for increasing  delays (see also Fig.$\,$2a), while the
line width does not change appreciably (see also Fig.$\,$3a).

For $i=45\deg$ (Fig.$\,$4b) the projected velocities are higher and produce
a broadening of the 2D-TF with respect to the low inclination case.
The maximum of the 1D-TF corresponds to the edge of the internal ellipse
that can be clearly seen in Fig.$\,$4b, where a high value of the 2D-TF is
achieved for blueshifted energies. The high intensity of the 2D-TF in the inner
edge of the ellipse affects also the centroid energy, which achieves its
maximum slightly before $\tau_3$. The line width maximum is reached around
$\tau_3$ as a result of the contributions from the red part of the 2D-TF.

For $i=80\deg$ (Fig.$\,$4c) an even stronger blue component of  the 2D-TF
develops, corresponding to the highest value of $E_c$ and of  the line width
for $\tau\sim\tau_3$ (note that the maximum of $E_c$ is for  times slightly
smaller than $\tau_3$). This effect derives from gravitational light bending
and surface projection effects.

In the weak field limit ($r_i=50\,r_g$) the 2D-TF is still asymmetric  (i.e. it
is redshifted for every inclination for low inclinations and shows the
characteristic enhancement in the blue horn near $\tau_3$), but with a much
smaller amplitude.

\subsection{Bulge-like source}

Several lines of evidence support the view that the dimensions of the
central illuminating region in AGNs are not negligible in comparison with
the emitting disk regions.
Chen, Halpern \& Filippenko (1989) have shown that reprocessing
in a standard geometrically thin disk does not reproduce the properties of
the $H\alpha$ and $H\beta$ lines in Arp 102B.
Molendi et al. (1992) have shown that a modified self-irradiating disk,
in which the central region is blown up into a bulge-like shape,
can account for the optical and UV variability of NGC~5548.
Finally, the strength of the fluorescence iron K$\alpha$ lines in many
AGNs (EW $\sim 100-200$~eV), requires that the accreting material
subtends larger solid angles than the thin disk model
allows (George \& Fabian 1991; Matt et al. 1991).
A possible solution to these problems is represented by a
standard disk, the innermost regions of which have become geometrically
thick possibly as a consequence of radiation-pressure related instabilities
(e.g. Shapiro, Lightman \& Eardley 1976; Wandel \& Liang 1991).
In order to investigate the response of the disk line emissivity to
variations from an extended central source, we consider the limiting case
of an optically thin
spherical source of radius $r_c$ that extends up to the disk inner radius
(i.e. $r_c=r_i$). The emissivity law for this geometry,
$\epsilon(r)=r^{q(r)}$, is calculated following the method outlined in
Chen \& Halpern (1989) (note that photons from the central source are
assumed to propagate along straight lines).
The value of $q(r)$ is $<-3$ for small radii and approaches $q\sim -3$ for
larger radii ($r\gsim 3\,r_i$, see Fig.$\,$5b in Chen \& Halpern 1989).
The fraction of radiation emitted by the central source which is
intercepted by the disk is $\sim 25\%$ (Chen \& Halpern 1989), about half
than in the case of a point-like source above the disk plane as discussed
by Matt \& Perola (1992).

To calculate the line response a prescription for the fastest possible
variation of the extended central source is also needed: we assume that a
point in the volume of the central source varies its emission
simultaneously with that of any other point.
The intensity of the continuum variation that reaches a point in the disk
depends on its distance, due to the different solid angle under which the
central sphere is seen.
In particular, the continuum intensity that reaches at a given time
delay $\bar\tau$ (from the turning on of the central source) a point
located at $(r,\,\phi)$ must be proportional to the area of the
intersection between the central source and the sphere centred at ($r,\,
\phi$) of radius $\bar\tau\,c$.
With this geometrical prescription we derive the intensity that an
observer at infinity sees at a time delay $\tau=\bar\tau\,
(1-\sin\phi\,\sin i)$ (see Eq.$\,$3), reaching the considered point:
$$
I_c(r,\,\tau)={{I_c^0\,\tau}\over{{r_c}^2}}\,\bigl(\tau
-{{{\tau}^2+r^2-{r_c}^2}\over{2\,r}}\bigl) \eqno(5)
$$
\noindent with $I_c^0$ the value of the maximum intensity.

For this geometry, the echo front has a finite thickness.
Its boundaries, neglecting photon bending, are
$\tau_{in}={{r-r_{c}}\over{c}}\,(1-\sin i \sin \phi)$ and
$\tau_{out}={{r+r_{c}}\over{c}}\,(1-\sin i \sin \phi)$,
which are ellipses with the most distant focus in the source
equator at $\phi=270\deg$ and $\phi=90\deg$ respectively.
The evolution of the echo front is characterised by two times:
the first corresponds to the inner part of the echo
front entering the disk ($\tau^{ext}_1=0$); the second to the outer part of
the echo front being fully included in the line emitting part of the
disk ($\tau^{ext}_3=2\,r_i/c\,(1+\sin i)$).

\subsubsection{Line moments evolution}

Strictly speaking, TFs cannot be calculated for this geometry, as, by
definition, they give the response to a $\delta-$like continuum variation.
We calculate instead the response of the line moments to the assumed
(shortest) variation of the central source.
The evolution of the line intensity and the higher line moments is computed
for the same set of parameters as in the limit of a point-like source limit.
Note that, in all cases, the line response is instantaneous because the
disk is assumed to straddle the equator of the central source (i.e. $r_i=r_c$).

The line intensity increases, giving rise to a broad maximum for time
delays of $2-3\,r_c/c$ depending on inclination (see Fig.$\,$5).
This broad peak is clearly suggestive of the convolution of the 1D-TF for
the point-like source model (see Fig.$\,$1) with the assumed variation of
the extended source (note however that the radial dependence of the line
emissivity is somewhat different in the two models).

Similarly, the line centroid energy does not show rapid variations (Fig.$\,$6),
but retains the general properties described for the case of a
point-like central source: depending on inclination, it increases
monotonically to unity or gives rise to a single broad maximum and then
decays ($i\gsim 40\deg$).

The line width shows a rapid rise to a {\it plateau} followed by a slow
decay (Fig.$\,$7).
This {\it plateau} is characterised by a broad peak which is less
pronounced for low inclinations in the weak field (Fig.$\,$7b).
Note that in the strong field case the line width peaks at higher
time delays for increasing inclinations (Fig.$\,$7a),
whereas the opposite trend is apparent in the weak field case (Fig.$\,$7b).
This phenomenon clearly represent the equivalent of the characteristic
feature of Fig.$\,$3a.

In general, due to the longer intensity variation at the
central source, the evolution of the line moments in the case of an
extended central source presents with considerably smoother features
compared to that of a point-like central source and, in particular, all
sharp features are washed out.

\section{DISCUSSION}

In this paper, by extending previous works, we have calculated the response
of a relativistic accretion disk to a continuum variation from a central
illuminating source. We have considered two limiting geometries: a point-like
central source and an extended spherical source straddling the inner edge of
the disk. Realistic geometries are likely to be in between these two extreme
cases.
We have identified a number of characteristic features, which could help
assessing the presence of an accretion disk and evaluating its parameters.
If the central illuminating source is point-like and lies in the disk plane,
a disk geometry can be recognised through some features of the response
function which are {\it qualitatively} the same for relativistic and
non-relativistic disks.

For a relativistic accretion disk the strength of the gravitational field in
the vicinity of the central object introduces substantial changes in the
characteristics of the response functions. In particular, for a point-like
central source the second peak of the 1D-TF dominates for every disk
inclination (Fig.$\,$1a). The evolution of the line centroid energy is fully
dominated by relativistic effects, showing a monotonic rise to the line rest
energy for low inclinations and a broad blueshifted maximum for high
inclinations (Fig.$\,$2a).
The response of the line width is affected by gravitational lensing and
surface projection effects; its peaks corresponds to the time in which the
whole echo front is within the line emitting region of the disk ($\tau_3$).
For high inclinations these effects become very evident (Fig.$\,$3a; see also
Matt \& Perola 1992; Matt, Perola \& Stella 1993).

For a spherical central source extending to the innermost disk regions, the
stationary line profile remains basically unchanged, while the sharp features
of the line intensity response and of higher line moments are smoothed out by
the longer duration of the central source variations. The intensity response
function shows a broad maximum for any inclination and disk inner radius. The
evolution of the line centroid energy and width resembles that for a
point-like central source, even if the sharp features are smoothed out.

The response of the line moments can be especially useful to derive geometric
information in those cases in which the energy resolution of the detector is
insufficient to study the detailed line profile.
In principle, the reverberation technique allows not only to infer disk
geometry but also to estimate disk parameters, by studying the different
behaviour of the line intensity response function and higher line moments or,
alternatively, the 2D-TF.  Moreover, in the case in which the presence of an
accretion disk and its parameters are inferred based on the stationary line
profile, the reverberation properties can be used as a consistency check
(Mannucci et al. 1992).

In Campana \& Stella (1993) we calculated the stationary values of the line
centroid energy and width for a relativistic Keplerian accretion disk with an
extended central source. Clearly the information of these stationary values
is not sufficient to constrain univocally the disk parameters.
To further constrain these parameters and derive an absolute scale length
related to the light propagation time delays, different approaches can be
adopted. One of them consists in selecting the values of the line centroid
energy and width corresponding to the time delay at which the response
of the line intensity reaches its maximum ($\tau_{max}$). The dependence of
$E_c(\tau_{max})$ and $W(\tau_{max})$ on the disk parameters $r_i$ and $i$ is
stronger than the stationary values of $E_c$ and $W$ (compare Fig.$\,$8
with Fig.$\,$1 in Campana \& Stella 1993).
Alternatively the maximum and/or minimum values of the response of the line
centroid energy and the maximum of the response of the line width can be
considered. Note however that these values are somewhat dependent on the
geometry assumed for the central source (compare Figs.$\,$2 and 3 with
Figs.$\,$6 and 7).
The latter prescription is similar to the one proposed by Matt \& Perola
(1992).
These methods allow to constrain the disk parameters and derive the mass of
the central object by combining the absolute time measurement of $\tau_{max}$
(or the delay for which the maximum of $E_c(t)$ and/or $W(t)$ is reached)
with the relative measurement in units of $r_g/c$ inferred from the line
response or the characteristics of the stationary line profile.

\bigskip
{\it Acknowledgemnets}

\smallskip
{\noindent We acknowledge useful comments by an anonymous referee.}

\bigskip

\centerline{\bf FIGURE CAPTIONS}

{\bigskip
\hangindent=.6cm \hangafter=1
\noindent
{\bf Fig.$\,$1.}
Line intensity response of a relativistic Keplerian accretion disk to a
short rectangular flux variation produced at a point-like central source
as a function of time delay in units of
$r_g/c=G\,M/c^3=4.9\times10^{-6}M/\!\msole$~s.
The first panel (a) refers to an inner radius $r_i=6\,r_g$,
while the second (b) to $r_i=50\,r_g$;
different curves correspond to different inclinations.
All curves are normalised to a  maximum value of 1.
Here and in Figs.$\,$2 and 3 the characteristic delays discussed in the
text ($\tau_1$, $\tau'_2$ and $\tau_3$) are indicated with different
symbols.
\par}

{\smallskip
\hangindent=.6cm \hangafter=1
\noindent
{\bf Fig.$\,$2.}
Response of the line centroid energy (normalised to the rest energy)
of a relativistic disk to a short rectangular flux variation
at a point-like central source as a function of time delay.
Inner radii and inclinations are as in Fig.$\,$1.
\par}

{\smallskip
\hangindent=.6cm \hangafter=1
\noindent
{\bf Fig.$\,$3.}
Response of the line width (normalised to the rest energy) of a
relativistic disk to a short rectangular flux variation at a point-like
central source as a function of the time delay.
Inner radii and inclinations are as in Fig.$\,$1.
\par}

{\smallskip
\hangindent=.6cm \hangafter=1
\noindent
{\bf Fig.$\,$4.} 2D-TF for a relativistic accretion
disk orbiting a point-like central source.
The disk inner radius is $r_i=6\,r_g$.
Panel (a) shows the 2D-TF for an inclination angle $i=10\deg$,
panel (b) for $i=45\deg$ and panel (c) for $i=80\deg$.
\par}

{\smallskip
\hangindent=.6cm \hangafter=1
\noindent
{\bf Fig.$\,$5.}
Line intensity response of a relativistic accretion disk
to a short variation produced by an extended source (see text)
as a function of time delay $\tau$ in units of $r_g/c$.
Panel (a) refers to $r_i=r_c=6\,r_g$, panel (b) to $r_i=r_c=50\,r_g$.
Different curves correspond to different inclinations.
\par}

{\smallskip
\hangindent=.6cm \hangafter=1
\noindent
{\bf Fig.$\,$6.}
Response of the line centroid energy (normalised to the rest energy)
of a relativistic disk to a short variation of an extended
central source as a function of the time delay.
Inner radii and inclinations are as in Fig.$\,$5.
\par}

{\smallskip
\hangindent=.6cm \hangafter=1
\noindent
{\bf Fig.$\,$7.}
Response of the line width (normalised to the rest energy) of a
relativistic disk to a short variation of an extended
central source as a function of the time delay.
Inner radii and inclinations are as in Fig.$\,$5.
\par}

{\smallskip
\hangindent=.6cm \hangafter=1
\noindent
{\bf Fig.$\,$8.}
Normalised line centroid energy (panel a) and normalised
line width (panel b) versus inclination for a time delay corresponding to the
maximum of the line intensity response, for an extended central source.
Different curves correspond to dif\-ferent inner radii ($r_i= 6, 8, 10,
12.5, 15, 17.5, 20, 25 ,30, 40, 50, 100$ $r_g$). Panel c gives the time
delay (in units of $r_g/c$) corresponding to the maximum of the line
intensity versus inclination.
\par}

\bsp

\label{lastpage}


\begin{thebibliography}{}

\bibitem{ar}
Arnaud, K.A. et al.,
1985, \mon 217 105 \par


\bibitem {ba}
Barr, P., White, N.E. \& Page, C.G., 1985 , \mon 216 65P \par


\bibitem {bl}
Blandford, R.D. \& McKee, C.F., 1982, \apj 255 419


\bibitem {cs93}
Campana, S. \& Stella, L., 1993, \mon 264 395


\bibitem {cl90}
Clavel, J. et al.,
1990, \mon 246 668


\bibitem {cl91}
Clavel, J. et al., 1991, \apj 366 64


\bibitem {chf}
Chen, K., Halpern, J.P. \& Filippenko, A.V., 1989, \apj 339 742


\bibitem {ch}
Chen, K. \& Halpern, J.P., 1989, \apj 344 115


\bibitem {d}
Day, C.S.R., Fabian, A.C., George, I.M. \& Kunieda, H., 1990, \mon 247 15P


\bibitem {eh}
Eracleous, M. \& Halpern, J.P., 1994, \apjs 90 1


\bibitem {er}
Eracleous, M., Livio, M., Halpern, J.P. \& Storchi-Bergmann, T., 1995, ApJ,
10 January 1995 issue


\bibitem {fa}
Fabian, A.C. et al., 1994, \pasj 46 L59


\bibitem {frsw}
Fabian, A.C., Rees, M.J., Stella, L. \& White, N.E., 1989,
\mon 238 729


\bibitem {gf}
George, I.M. \& Fabian, A.C., 1991, \mon 249 352


\bibitem {h}
Horne, K., 1985, \mon 213 129


\bibitem {hwp}
Horne, K., Welsh, W.F. \& Peterson, B.M., 1991, \apj 367 L5


\bibitem {kw}
Kallman, T. \& White, N.E., 1989, \apj 341 955


\bibitem {kh}
Krolik, J.H., Horne, K., Kallman, T.R., Malkan, M.A.,
Edelson, R.A. \& Kriss, G.A., 1991, \apj 371 541


\bibitem {ms}
Malkan, M.A. \& Sargent, W.L.W., 1982, \apj 254 22


\bibitem {mss}
Mannucci, F., Salvati, M. \& Stanga, R.M., 1992, \apj 394 98


\bibitem {mz}
Maoz, D. et al.,
1991, \apj 367 493


\bibitem {mrw}
Mardaljevic, J., Raine, D.J. \& Walsh, D., 1988, Ap. Lett. Comm. 26 357


\bibitem {mh88}
Marsh, T.M. \& Horne, K., 1988, \mon 235 269


\bibitem {mh90}
Marsh, T.M. \& Horne, K., 1990, \apj 349 593


\bibitem {mfr}
Matt, G., Fabian, A.C. \& Ross, R.R., 1993, \mon 262 179


\bibitem {mp}
Matt, G. \&  Perola, G.C., 1992, \mon 259 433


\bibitem {mpp}
Matt, G., Perola, G.C. \& Piro, L., 1991, \aa 247 25


\bibitem {mps}
Matt, G., Perola, G.C. \& Stella, L., 1993, \aa 267 643


\bibitem {mol}
Molendi, S., Maraschi, L. \& Stella, L., 1992, \mon 255 56


\bibitem {na}
Nandra, K. \& Pounds, K.A., 1994, MNRAS in press


\bibitem {pz}
P\'erez, E., Robinson, A. \& de la Fuente, L., 1992, \mon 256 103


\bibitem {pet}
Peterson B.M., 1993, \pasp 105 247


\bibitem {rb}
Rouchi, K.P. \& Blandford, R.D., 1994, \apj 421 46


\bibitem {sh}
Shapiro, S.L., Lightman, A.P. \& Eardley, D.M., 1976, \apj 204 187


\bibitem {sk}
Skilling, J. \& Bryan, R.K., 1984, \mon 211 111


\bibitem {sm}
Smale, A.P. et al., 1993, \apj 410 796


\bibitem {st}
Stella, L., 1990, \nat 344 747


\bibitem {sc}
Stella, L. \& Campana, S., 1991, in {\sl Iron Line Diagnostic in X-ray
Sources}, Eds. Treves, A., Perola, G.C. \& Stella, L. (Springer-Verlag), 230


\bibitem {u}
Ulrich, M.-H. et al., 1991, \apj 382 483


\bibitem {wa}
Wandel, A. \& Liang, E.P., 1991, \apj 380 84


\bibitem {we}
Welsh, W.F. \& Horne, K., 1991, \apj 379 586


\bibitem {z}
Zheng, W., Veilleux, S. \& Grandi, S.A., 1991, \apj 381 418

\end{thebibliography}
\end{document}